# Quantum Stochastic Gradient Descent in its continuous-time limit based on the Wigner formulation of Open Quantum Systems

J.A. Morales Escalante[a,b,]

[a]*Department of Mathematics, FLN Building, University of Texas at San Antonio, San Antonio, Texas, USA 78249*
[b]*Department of Physics and Astronomy, AET Building, University of Texas at San Antonio, San Antonio, Texas, USA 78249*

**Abstract**

The main ideas behind a research plan to use the Wigner formulation as a bridge between classical and quantum probabilistic algorithms are presented, focusing on one particular case: the Quantum analog of Stochastic Gradient Descent in its continuous-time limit based on the Wigner formulation of Open Quantum Systems.

*Keywords:* Wigner formulation of Open Quantum Systems, Continuous time limit, Stochastic Gradient Descent

## 1. Introduction

The Wigner formalism of quantum mechanics presents a completely quantum mechanical phase space description, at the cost of losing the non-negativity of probability density functions from which observables will be calculated through integrals over space. The Wigner picture is very intuitive, as it presents itself as a quantum extension of typical classical statistical mechanics descriptions of probability distributions in phase space linked to Boltzmann equations and other families of kinetic equations as well. This Wigner quantum analog of a classical probability density function over phase space shows the similarities but also stresses the differences between the quantum and classical worlds (sometimes behaving as a false cognate if taken naively without care or physical intuition). This intuitive physical picture of the Wigner function renders itself helpful in the study of open quantum systems. Under the assumption of a Markovian noise, the density matrix picture for these quantum subsystems perturbed by their environment is given by the Lindblad master equation, which has the main drawback that the operators representing the noise effect coming from the environment are somehow obscure and hard to interpret in a physically intuitive way even though their math structure is well understood from the operator algebra point of view. On the other hand, the completely equivalent Wigner-Fokker-Planck equation (in continuous variables under the assumption that a phase space description is possible) is a more amenable and physically intuitive mathematical model. Under the assumption of continuous variables, the density matrix description of Markovian open quantum systems under a Lindblad master equation is completely equivalent to the phase-space formulation of open quantum systems via a Wigner-Fokker-Planck equation, since one can obtain the Wigner quasi-probability density function as the Fourier transform of the density matrix in convenient coordinates (and revert back from the Wigner function to the density matrix expressed in a position basis via an inverse Fourier transform). If one assumes that the potential is harmonic as in $V = x^2/2$, the Wigner-Fokker-Planck equation takes the form (taking units such that $\hbar = 1 = m$)

$$w_t + p \cdot w_x - x \cdot w_p = \nabla \cdot (D \nabla w) + \gamma \nabla_p \cdot (wp) \quad (1)$$

where $D$ is the diffusion matrix representing part of the noise process introduced by the environment, and $\gamma$ is a friction coefficient for the other term introducing noise. Though this benchmark problem is already very useful since a good amount of the potentials that physicists use to model phenomena are linear combinations of harmonic potentials, one could ask the question on how to use a Wigner picture for more general non-harmonic potentials. The Wigner-Fokker-Planck equation for an arbitrary potential $V$ is given by

$$w_t + p \cdot w_x + \Theta[V](w) = \nabla \cdot (D \nabla w) + \gamma \nabla_p \cdot (wp)$$
$$\Theta[V](w) = \frac{-i}{\hbar(2\pi)^d} \int_{\mathbb{R}^{2d}} \delta V(x, \eta) w(x, p') e^{i\eta \cdot (p-p')} dp' d\eta \quad (2)$$

where $\delta V(x, \eta) = V(x + \frac{\eta}{2}) - V(x - \frac{\eta}{2})$. The right-hand-side terms have a clear interpretation as first a diffusion operator and then a friction term (that can be treated as a transport). In the left hand side, there is quantum transport over the phase space, with the pseudo-differential (integral) operator $\Theta[V](w)$. This operator is the result of the Fourier-transform of the commutator $[\widehat{\rho}, V]$ in the Lindblad picture for density matrices $\widehat{\rho}$. As we have mentioned above, the relation between the density matrix operator $\widehat{\rho}$ in its position representation and the Wigner function is given by a Fourier transform in convenient coordinates, given below,

$$w(x, p, t) = \frac{1}{\hbar(2\pi)^d} \int_{\mathbb{R}^{2d}} \rho(x+y, x-y) e^{-iy \cdot p/\hbar} dy$$
$$\rho(x, y) = \langle x+y | \widehat{\rho} | x-y \rangle. \quad (3)$$

[*]Corresponding author
 *Email address:* jose.morales4@utsa.edu (J.A. Morales Escalante)
 *URL:* http://josemoralesescalante.wordpress.com (J.A. Morales Escalante)

## 1.1. Stochastic Gradient Descent

We focus on the study of stochastic gradient descent by means of their continuous time limit (Su, Wibisono, Jordan, Shi). Following (Liu, Su, Li), if $f : \mathbb{R}^d \to \mathbb{R}^d$ is the objective function constructed with all data, and the Stochastic Gradient Descent (SGD) algorithm is given by

$$x_{k+1} = x_k - s\widetilde{\nabla} f(x_k)$$

where $s$ is the learning rate and $\widetilde{\nabla} f$ is the estimated gradient (obtained from a mini-batch), then one can model SGD by the following iteration,

$$x_{k+1} = x_k - s\nabla f(x_k) - s\xi_k$$

where $\xi_k$ is a given noise that obeys a normal distribution, as in the so-called "unadjusted Langevin dynamics". Taking the continuous time limit one obtains a stochastic differential equation (SDE) that includes the learning rate $s$ as a parameter, which models the discrete SGD algorithm. This SDE is

$$dx = -\nabla f(x)dt + \sqrt{s}dW,$$

with $W$ a Brownian motion.

There is a direct analogy with open quantum systems via this SDE. In the case where the environment introduces Markovian noise, the SDE version of the Wigner-Fokker-Planck model of open quantum systems is given by an SDE modeling the evolution of the Wigner quasi-distribution $w(x, p, t)$. Taking units such that $\hbar = 1 = m$, if we consider $\alpha$ the quantum transport over the subsystem, $\sigma$ the standard deviation related to the noise, $2\sigma dW = 2Ddt\mathcal{N}(0, 1)$ where $\mathcal{N}(0, 1)$ is a normal distribution with mean zero and variance 1, and $D$ the diffusion matrix representing part of the noise introduced by the environment. The other part of the noise is given by a friction term which is included in the quantum transport. Namely, the friction term is $\gamma \nabla_p \cdot (wp)$, where $\gamma$ is the friction coefficient and $p$ is the momentum variable. The full transport term (including both pure quantum transport and the friction term representing noise but added to the transport term) is

$$\alpha(x, p, t) = \nabla \cdot (w(p, -x - \gamma p)).$$

From a particle viewpoint, where $z = (x, p)$ represents the location in phase space of particle points to be sampled from a Wigner quasi-PDF, following D'Ariano et al., one can represent the motion of phase-space points according to the following model,

$$dz_i = Q[z_i(t)]dt + E[z_i(t)]$$

with $i \in \{1, N\}$ an index representing the studied particle from the sample, $Q[z(t)] = (p, -x - \gamma p)$ being the quantum transport, and $E$ being a zero average Gaussian random variable, whose variance is $Ddt$. The closest expression as a stochastic gradient for our SDE related to open quantum systems is via the Hamiltonian+gradient flow as in

$$dz_i = [J\nabla_i H(z) - \nabla_i(\gamma p^2/2)]dt + E[z_i(t)], \quad i \in \{1, \cdots, N\}$$

where part of the flow is indeed gradient (the friction one) but the other part is related to a Hamiltonian flow where the gradient is multiplied by the symplectic matrix $J$. Our problem has, regarding transport, both a symplectic flow structure (due to the Hamiltonian) and a gradient flow structure too (due to the friction term), which combined with the diffusion due to the environment noise, has a dynamics almost similar to the SGD counterpart (becoming more similar to a gradient flow as the friction term dominates in the limit $\gamma \to \infty$). To fully express our problem then, we have then that

$$d(x_i, p_i) = [J\nabla_i(\frac{p^2 + x^2}{2}) - \gamma \nabla_i(\frac{p^2}{2})]dt + E[x_i(t), p_i(t)], \quad i \in \{1, \cdots, N\}$$

Our point: the Hamiltonian flow (for the case of the harmonic oscillator dynamics) plus a dominant gradient-flow (in the case where the friction dominates by making the damping coefficient $\gamma \to \infty$) and diffusion terms (the last two due to subsystem-environment interaction) is almost equivalent to the continous limit of Stochastic Gradient Descent (when friction dominates Hamiltonian transport), and therefore their behaviors (and performances) are similar (the similarity becoming stronger as friction dominates more the Hamiltonian transport). Since the numerics of SGD is similar to the Physics of open quantum systems, we would expect variations of SGD to a good numerical model available for mimicking quantum problems where the hamiltonian transport is subjected to environment noise represented in both transport due to friction and diffusion connected to random walks happening continuously in time.

**Conjecture 1.** *For minimization problems that can be formulated as quantum transport perturbed by Markovian noise, then SGD must a good candidate for a numerical method available, since it's reproducing exactly the problem it's trying to solve.*

*1.1.1. Advantages of an SGD formulation as an open quantum system rather than a closed quantum dynamics*

A closed quantum system description has the drawback that their evolution is unitary, so a Quantum Tunneling Walk (QTW) will not formally converge. This is not the behavior observed for SGD, whose dynamics is quite similar to an open quantum system: gradient-related transport plus noise, and convergence to a steady state solution (as proven for the benchmark problem of a harmonic potential). So an open quantum system dynamics does have truly a mixing time as exponential decay towards the steady state solution, for example in the case of a harmonic potential, where it has been proved by Sparber et al [6].

## 2. Wigner-Fokker-Planck: harmonic case in arbitrary dimension, decay rate, mixing time

In classical machine learning, it is known [1],[2],[3],[8] that the continuous-time limit of gradient descent methods such as the Stochastic Gradient Descent (SGD) algorithm can be modeled by differential equations. For SGD, whose Stochastic Differential Equation (SDE) approximation is

$$dx = -\nabla f(x)dt + \sqrt{s}dW,$$

with $W$ a standard Brownian motion, the Fokker–Planck equation for the SGD pdf $\rho_{\text{SGD}}$,

$$\partial_t \rho_{\text{SGD}} = \nabla \cdot (\rho_{\text{SGD}} \nabla f) + \frac{s}{2} \nabla^2 \rho_{\text{SGD}},$$

is the respective continuous-time limit equivalent [14],[12],[13].

Quantum extensions of SGD have been proposed attempting to obtain a quantum advantage. Work in [14] explores quantum speedups for nonconvex optimization problems introducing a quantum tunneling walk (QTW) algorithm for nonconvex problems where local minima are "approximately global". This is obtained by analogy between classical Fokker-Planck and the Schroedinger equation as the quantum equivalent under certain periodicity conditions. They report quantum speedup over classical stochastic gradient descents (SGD) when the barriers between different local minima are high but thin and flat minima.

However, it is physically clear that the quantum version of the equation representing the SDE version of SGD should be Wigner-Fokker-Planck, which for the case of a harmonic potential $V(x) = \omega_0^2 x^2/2$ takes the form

$$\partial_t w = -\nabla \cdot (w J \nabla H) + \gamma \nabla_p \cdot (w \nabla_p \frac{p^2}{2}) + \nabla \cdot (D \nabla w), \quad J = \begin{pmatrix} 0 & I_d \\ -I_d & 0 \end{pmatrix}, \quad (4)$$

with the Hamiltonian $H = T(p) + V(x) = \frac{p^2}{2m} + \frac{\omega_0^2 x^2}{2}$ representing the (kinetic plus potential) energy, $J$ the symplectic matrix, $d$ the dimension of the physical space, the diffusion matrix $D$ and friction coefficient $\gamma$, because the open quantum system model represented above is the quantum version of the system studied by the classical Fokker-Planck representing an SGD, particularly when the gradient term associated to friction dominates the Hamiltonian transport. Wigner-Fokker-Planck is completely equivalent to the dynamics of a noisy quantum transport system whose SDE form is

$$d(x,p) = (J \nabla H - \nabla_p \frac{\gamma p^2}{2}) dt + E(D), \quad (5)$$

where $x$ is the position and $p$ the momentum in quantum phase space, and $E$ is a random variable sampled from a Gaussian distribution with covariance matrix $2Ddt$ ($dt \approx \Delta t$ discretized). We illustrate in Figure 1 below the dynamics of the Wigner-Fokker-Planck model for an open quantum system as the balance between quantum transport (made of both a Hamiltonian flow and gradient transport) and diffusive noise.

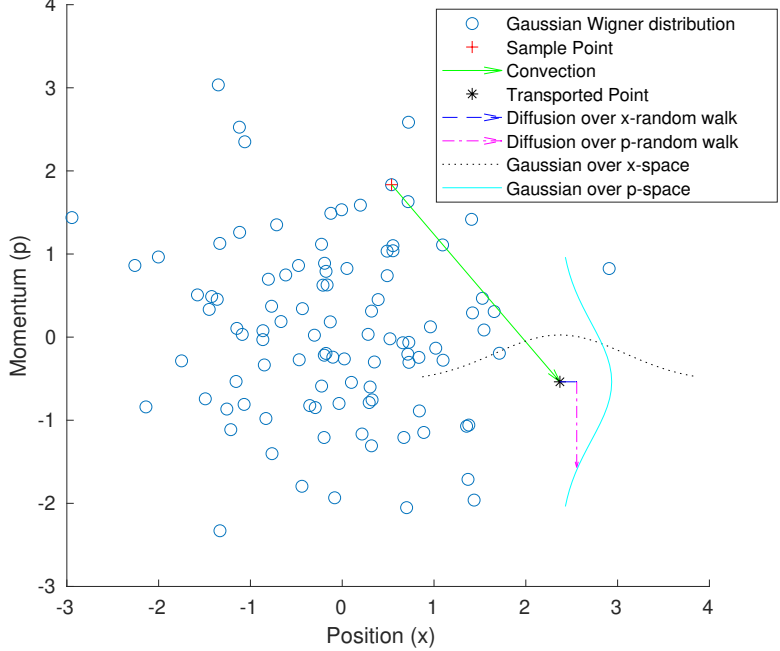

Figure 1: A Wigner distribution is shown, where quantum transport & diffusion mechanisms over a point are highlighted. This process is the continuous quantum limit of Stochastic Gradient Descent.

There are previous studies of the harmonic benchmark for WFP [6], whose results can be used. In particular, Sparber et al. prove in [6] the existence of a steady-state solution $w_\infty(x,p)$ for this benchmark problem, providing its analytical form, and it also gives the exponential decay rate of any well-posed initial condition towards this analytical steady state for the harmonic case. Namely, [6] finds that

$$||w(t,\cdot) - w_\infty||_p \leq C\exp(-\kappa t), \quad C > 0, t > 0,$$

with $1 \leq p \leq 2$ depending on the relative entropy (logarithmic or quadratic), and that the respective decay rate is the smallest eigenvalue of the Hessian of the quadratic polynomial $A$ such that $w_\infty \propto \exp(-A(x,p))$ (the Wigner steady state for this problem is a Gaussian state), so that

$$\text{Hess}\, A \geq \kappa I_{2d}, \quad (\text{Hess}\, A)_{ij} = \partial^2_{ij} A, \quad i,j = 1,\cdots,2d.$$

This exponential decay rate is advantageous for interpreting WFP as a quantum extension of SGD because it will provide the respective mixing time of a quantum SGD algorithm based on the WFP equation. Most importantly, nothing precludes interpreting WFP not only on physical dimensions such that $d = 1, 2, 3$, but actually for $d >> 1$ as is typically the case for Machine Learning, in which the problem is of high dimensionality given the amount of data for the problem. Our result is that, for WFP in a high dimension $d \gg 1$, the decay rate providing the mixing time for the related quantum SGD algorithm in the benchmark case of a harmonic potential is simply the smallest of only 2 possible roots (with high algebraic multiplicity) of the characteristic polynomial, which although of degree $2d$, as we mentioned has only 2 possibly different roots of multiplicity $d$.

Our preliminary result indicates that the decay rate towards the steady state in WFP (and therefore the mixing time in its quantum version of SGD) does not depend on the dimension $d$ of the problem for the harmonic potential benchmark problem, as it should be since this doesn't happen either in the classical SGD optimization for a quadratic function (such as the parabola representing a harmonic potential). Namely, the decay rate does not slow down as the quantum space dimension $d \gg 1$ increases for this benchmark case of a harmonic potential.

## 2.1. Preliminary Calculations

We show the preliminary calculations for high dimensions $d \gg 1$. We consider $x \in \mathbb{R}^d$, $p \in \mathbb{R}^d$, so $(x,p) \in \mathbb{R}^{2d}$, and therefore the diffusion matrix is the block matrix

$$D = \begin{pmatrix} D_{qq}I_d & D_{pq}I_d \\ D_{pq}I_d & D_{pp}I_d \end{pmatrix}, \quad D_{pq} = D_{qp},$$

where $D_{ij} \in \mathbb{R}^{d \times d}, i,j = q,p$. As mentioned by Sparber et al. [6], the optimal exponential rate $\kappa$ is the smalles eigenvalue of the Hessian of $\widetilde{A}$. The potential $\widetilde{A}$ is defined by

$$\widetilde{A}(\widetilde{y}) = A(\sqrt{D}y^T),$$

with $\sqrt{D}$ the square root of $D$ in the sense of positive definite matrices, and $y = (x,p)$.

We know the form of $A$ for the steady state, given by

$$A(x,p) = \frac{\gamma}{Q}[Q_{11}\omega_0^2 x^2 + 2Q_{12}\omega_0 x \cdot p + Q_{22}p^2]$$

$$Q_{11} = D_{pp} + \omega_0^2 D_{qq}$$

$$Q_{12} = 2\omega_0 \gamma D_{qq}$$

$$Q_{22} = D_{pp} + \omega_0^2 D_{qq} + 4\gamma(D_{pq} + \gamma D_{qq}) = Q_{11} + 4\gamma(D_{pq} + \gamma D_{qq})$$

$$Q = Q_{11}Q_{22} - Q_{12}^2$$

Since to compute the Hessian of the modified potential $\widetilde{A}$ we do it against the variables $\widetilde{y}$, it is equivalent to compute the Hessian of $A$ against the variable $y$. So we simply consider

$$H = \text{Hessian}(A) = \frac{\gamma}{Q}\begin{pmatrix} Q_{11}\omega_0^2 & Q_{12}\omega_0 \\ Q_{12}\omega_0 & Q_{22} \end{pmatrix}$$

$$H = \frac{\gamma}{Q} \begin{pmatrix} [D_{pp} + \omega_0^2 D_{qq}]\omega_0^2 & 2\omega_0^2 \gamma D_{qq} \\ 2\omega_0^2 \gamma D_{qq} & D_{pp} + \omega_0^2 D_{qq} + 4\gamma(D_{pq} + \gamma D_{qq}) \end{pmatrix}$$

with

$$Tr(H) = \frac{\gamma}{Q}(Q_{11}\omega_0^2 + Q_{22}) = \frac{\gamma}{Q}(Q_{11}\omega_0^2 + Q_{11} + 4\gamma(D_{pq} + \gamma D_{qq}))$$

$$Tr(H) = \frac{\gamma}{Q}([D_{pp} + \omega_0^2 D_{qq}]\omega_0^2 + D_{pp} + \omega_0^2 D_{qq} + 4\gamma(D_{pq} + \gamma D_{qq}))$$

$$Tr(H) = \frac{\gamma}{Q}([D_{pp} + \omega_0^2 D_{qq}](\omega_0^2 + 1) + 4\gamma(D_{pq} + \gamma D_{qq}))$$

and

$$\det(H) = (\frac{\gamma}{Q})^2(Q_{11}Q_{22} - Q_{12}^2)\omega_0^2 = (\frac{\gamma}{Q}\omega_0)^2 Q$$

So the characteristic polynomial for the Hessian of $A$ is, for $d = 1$,

$$\lambda^2 - Tr(H)\lambda + \det(H) = 0$$

$$\lambda^2 - \frac{\gamma}{Q}(Q_{11}\omega_0^2 + Q_{22})\lambda + (\frac{\gamma}{Q}\omega_0)^2(Q_{11}Q_{22} - Q_{12}^2) = 0$$

$$\lambda^2 - \frac{\gamma}{Q}(Q_{11}\omega_0^2 + Q_{22})\lambda + (\frac{\gamma}{Q}\omega_0)^2 Q = 0$$

therefore the roots are

$$\lambda_\pm = \frac{\frac{\gamma}{Q}(Q_{11}\omega_0^2 + Q_{22}) \pm \sqrt{[\frac{\gamma}{Q}(Q_{11}\omega_0^2 + Q_{22})]^2 - 4(\frac{\gamma}{Q}\omega_0)^2 Q}}{2}$$

$$\lambda_\pm = \frac{\frac{\gamma}{Q}(Q_{11}\omega_0^2 + Q_{22}) \pm \sqrt{[\frac{\gamma}{Q}(Q_{11}\omega_0^2 + Q_{22})]^2 - 4(\frac{\gamma}{Q}\omega_0)^2(Q_{11}Q_{22} - Q_{12}^2)}}{2}$$

$$\lambda_\pm = \frac{\frac{\gamma}{Q}(Q_{11}\omega_0^2 + Q_{22}) \pm |\frac{\gamma}{Q}|\sqrt{(Q_{11}\omega_0^2)^2 + 2Q_{22}Q_{11}\omega_0^2 + Q_{22}^2 - 4\omega_0^2(Q_{11}Q_{22} - Q_{12}^2)}}{2}$$

$$\lambda_\pm = \frac{\frac{\gamma}{Q}(Q_{11}\omega_0^2 + Q_{22}) \pm |\frac{\gamma}{Q}|\sqrt{(Q_{11}\omega_0^2)^2 - 2Q_{22}Q_{11}\omega_0^2 + Q_{22}^2 + 4\omega_0^2 Q_{12}^2}}{2}$$

$$\lambda_\pm = \frac{\frac{\gamma}{Q}(Q_{11}\omega_0^2 + Q_{22}) \pm |\frac{\gamma}{Q}|\sqrt{(Q_{11}\omega_0^2 - Q_{22})^2 + (2\omega_0 Q_{12})^2}}{2}$$

$$\lambda_\pm = \frac{\frac{\gamma}{Q}(Q_{11}\omega_0^2 + Q_{22}) \pm \sqrt{[\frac{\gamma}{Q}(Q_{11}\omega_0^2 - Q_{22})]^2 + (2\frac{\gamma}{Q}\omega_0 Q_{12})^2}}{2}$$

$$\lambda_\pm = \frac{\gamma(Q_{11}\omega_0^2 + Q_{22})}{2Q} \pm \sqrt{[\frac{\gamma}{2Q}(Q_{11}\omega_0^2 - Q_{22})]^2 + (\frac{\gamma}{Q}\omega_0 Q_{12})^2}$$

so the optimal exponential rate $\kappa$ is then the smallest eigenvalue,

$$\kappa = \frac{\gamma(Q_{11}\omega_0^2 + Q_{22})}{2Q} - \gamma\sqrt{(\frac{Q_{11}\omega_0^2 - Q_{22}}{2Q})^2 + (\frac{\omega_0 Q_{12}}{Q})^2}$$

then

$$\kappa = \frac{\gamma(Q_{11}\omega_0^2 + Q_{11} + 4\gamma(D_{pq} + \gamma D_{qq}))}{2Q} - \gamma\sqrt{[\frac{Q_{11}\omega_0^2 - Q_{11} - 4\gamma(D_{pq} + \gamma D_{qq})}{2Q}]^2 + (\frac{2\omega_0^2\gamma D_{qq}}{Q})^2}$$

$$\kappa = \frac{\gamma(Q_{11}[\omega_0^2 + 1] + 4\gamma[D_{pq} + \gamma D_{qq}])}{2Q} - \gamma\sqrt{[\frac{Q_{11}(\omega_0^2 - 1) - 4\gamma(D_{pq} + \gamma D_{qq})}{2Q}]^2 + (\frac{4\omega_0^2\gamma D_{qq}}{2Q})^2}$$

$$\kappa = \frac{\gamma(Q_{11}[\omega_0^2 + 1] + 4\gamma[D_{pq} + \gamma D_{qq}])}{2Q} - \gamma\sqrt{\frac{[Q_{11}(\omega_0^2 - 1) - 4\gamma(D_{pq} + \gamma D_{qq})]^2 + (4\omega_0^2\gamma D_{qq})^2}{(2Q)^2}}$$

$$\kappa = \frac{\gamma(Q_{11}[\omega_0^2 + 1] + 4\gamma[D_{pq} + \gamma D_{qq}])}{2Q}$$
$$-\gamma\sqrt{\frac{[Q_{11}(\omega_0^2 - 1)]^2 + [4\gamma(D_{pq} + \gamma D_{qq})]^2 - 8\gamma(D_{pq} + \gamma D_{qq})Q_{11}(\omega_0^2 - 1) + 16\omega_0^4\gamma^2 D_{qq}^2}{(2Q)^2}}$$

$$\kappa = \frac{\gamma(Q_{11}[\omega_0^2 + 1] + 4\gamma[D_{pq} + \gamma D_{qq}])}{2Q}$$
$$-\gamma\sqrt{\frac{[Q_{11}(\omega_0^2 - 1)]^2 + [4\gamma(D_{pq} + \gamma D_{qq})]^2 - 8\gamma Q_{11}(\omega_0^2 - 1)D_{pq} - 8\gamma^2 Q_{11}(\omega_0^2 - 1)D_{qq} + 16\omega_0^4\gamma^2 D_{qq}^2}{(2Q)^2}}$$

$$\kappa = \frac{\gamma(Q_{11}[\omega_0^2 + 1] + 4\gamma[D_{pq} + \gamma D_{qq}])}{2Q}$$
$$-\gamma\sqrt{\frac{[Q_{11}(\omega_0^2 - 1)]^2 + [4\gamma(D_{pq} + \gamma D_{qq})]^2 - 8\gamma Q_{11}(\omega_0^2 - 1)D_{pq} + 8\gamma^2 D_{qq}[2\omega_0^4 D_{qq} - Q_{11}(\omega_0^2 - 1)]}{(2Q)^2}}$$

and since
$$Q_{11} = D_{pp} + \omega_0^2 D_{qq}$$

$$\kappa = \frac{\gamma(Q_{11}[\omega_0^2 + 1] + 4\gamma[D_{pq} + \gamma D_{qq}])}{2Q}$$
$$-\gamma\sqrt{\frac{[Q_{11}(\omega_0^2 - 1)]^2 + [4\gamma(D_{pq} + \gamma D_{qq})]^2 - 8\gamma Q_{11}(\omega_0^2 - 1)D_{pq} + 8\gamma^2 D_{qq}[2\omega_0^4 D_{qq} - (D_{pp} + \omega_0^2 D_{qq})(\omega_0^2 - 1)]}{(2Q)^2}}$$

$$\kappa = \frac{\gamma(Q_{11}[\omega_0^2 + 1] + 4\gamma[D_{pq} + \gamma D_{qq}])}{2Q}$$
$$-\gamma\sqrt{\frac{[Q_{11}(\omega_0^2 - 1)]^2 + [4\gamma(D_{pq} + \gamma D_{qq})]^2 - 8\gamma Q_{11}(\omega_0^2 - 1)D_{pq} + 8\gamma^2 D_{qq}[2\omega_0^4 D_{qq} - \omega_0^2 D_{qq}(\omega_0^2 - 1) - D_{pp}(\omega_0^2 - 1)]}{(2Q)^2}}$$

$$\kappa = \frac{\gamma(Q_{11}[\omega_0^2 + 1] + 4\gamma[D_{pq} + \gamma D_{qq}])}{2Q}$$
$$-\gamma\sqrt{\frac{[Q_{11}(\omega_0^2 - 1)]^2 + [4\gamma(D_{pq} + \gamma D_{qq})]^2 - 8\gamma Q_{11}(\omega_0^2 - 1)D_{pq} + 8\gamma^2 D_{qq}[2\omega_0^4 D_{qq} - \omega_0^4 D_{qq} + \omega_0^2 D_{qq} - D_{pp}(\omega_0^2 - 1)]}{(2Q)^2}}$$

$$\kappa = \frac{\gamma(Q_{11}[\omega_0^2 + 1] + 4\gamma[D_{pq} + \gamma D_{qq}])}{2Q}$$

$$-\gamma\sqrt{\frac{[Q_{11}(\omega_0^2 - 1)]^2 + [4\gamma(D_{pq} + \gamma D_{qq})]^2 - 8\gamma Q_{11}(\omega_0^2 - 1)D_{pq} + 8\gamma^2 D_{qq}[\omega_0^4 D_{qq} + \omega_0^2 D_{qq} - D_{pp}(\omega_0^2 - 1)]}{(2Q)^2}}$$

$$\kappa = \frac{\gamma(Q_{11}[\omega_0^2 + 1] + 4\gamma[D_{pq} + \gamma D_{qq}])}{2Q}$$

$$-\gamma\sqrt{\frac{[Q_{11}(\omega_0^2 - 1)]^2 + [4\gamma(D_{pq} + \gamma D_{qq})]^2 - 8\gamma Q_{11}(\omega_0^2 - 1)D_{pq} + 8\gamma^2 D_{qq}[\omega_0^2 D_{qq}(\omega_0^2 + 1) - D_{pp}(\omega_0^2 - 1)]}{(2Q)^2}}$$

*2.1.1. Special case of unit frequency*

We consider the special case of $\omega_0 = 1$ for simplicity in the calculations. First we will do the $d = 1$ case and then we will consider $d >> 3 > 1$. In the special case considered, then

$$Q_{11} = D_{pp} + D_{qq}$$

$$Q_{12} = 2\gamma D_{qq}$$

$$Q_{22} = D_{pp} + D_{qq} + 4\gamma(D_{pq} + \gamma D_{qq}) = Q_{11} + 4\gamma(D_{pq} + \gamma D_{qq})$$

$$Q = Q_{11}Q_{22} - Q_{12}^2$$

$$Q = Q_{11}[Q_{11} + 4\gamma(D_{pq} + \gamma D_{qq})] - [2\gamma D_{qq}]^2$$

$$Q = Q_{11}^2 + 4\gamma(D_{pq} + \gamma D_{qq})Q_{11} - (2\gamma D_{qq})^2$$

$$Q = (D_{pp} + D_{qq})^2 + 4\gamma(D_{pq} + \gamma D_{qq})(D_{pp} + D_{qq}) - (2\gamma D_{qq})^2$$

$$Q = [(D_{pp} + D_{qq}) + 4\gamma(D_{pq} + \gamma D_{qq})](D_{pp} + D_{qq}) - (2\gamma D_{qq})^2$$

$$Q = [(D_{pp} + D_{qq}) + 4\gamma(D_{pq} + \gamma D_{qq})](D_{pp} + D_{qq}) - (2\gamma D_{qq})^2$$

In that case, since $\omega_0 = 1$,

$$\kappa = \frac{\gamma(Q_{11}[\omega_0^2 + 1] + 4\gamma[D_{pq} + \gamma D_{qq}])}{2Q}$$

$$-\gamma\sqrt{\frac{[Q_{11}(\omega_0^2 - 1)]^2 + [4\gamma(D_{pq} + \gamma D_{qq})]^2 - 8\gamma Q_{11}(\omega_0^2 - 1)D_{pq} + 8\gamma^2 D_{qq}[2\omega_0^4 D_{qq} - Q_{11}(\omega_0^2 - 1)]}{(2Q)^2}}$$

$$\kappa = \frac{\gamma(2Q_{11} + 4\gamma[D_{pq} + \gamma D_{qq}])}{2Q} - \gamma\sqrt{\frac{[4\gamma(D_{pq} + \gamma D_{qq})]^2 + 8\gamma^2 D_{qq} \times 2\omega_0^4 D_{qq}}{(2Q)^2}}$$

$$\kappa = \frac{\gamma(Q_{11} + 2\gamma[D_{pq} + \gamma D_{qq}])}{Q} - \gamma\sqrt{\frac{[2\gamma(D_{pq} + \gamma D_{qq})]^2 + 4\gamma^2 D_{qq}^2 \omega_0^4}{(Q)^2}}$$

$$\kappa = \frac{\gamma(Q_{11} + 2\gamma[D_{pq} + \gamma D_{qq}])}{Q} - \gamma\sqrt{\frac{[2\gamma(D_{pq} + \gamma D_{qq})]^2 + (2\gamma D_{qq}\omega_0^2)^2}{Q^2}}$$

$$\kappa = \frac{\gamma(Q_{11} + 2\gamma[D_{pq} + \gamma D_{qq}])}{Q} - 2\gamma^2 \sqrt{\frac{[(D_{pq} + \gamma D_{qq})]^2 + (D_{qq}\omega_0^2)^2}{Q^2}}$$

$$\kappa = \frac{\gamma(Q_{11} + 2\gamma[D_{pq} + \gamma D_{qq}])}{Q} - \frac{2\gamma^2}{|Q|}\sqrt{[(D_{pq} + \gamma D_{qq})]^2 + (D_{qq}\omega_0^2)^2}$$

$$\kappa = \frac{\gamma}{|Q|}[\text{sign}(Q)(Q_{11} + 2\gamma[D_{pq} + \gamma D_{qq}]) - 2\gamma\sqrt{[(D_{pq} + \gamma D_{qq})]^2 + (D_{qq}\omega_0^2)^2}]$$

so we have the proportionality relation

$$\kappa \propto [\text{sign}(Q)(Q_{11} + 2\gamma[D_{pq} + \gamma D_{qq}]) - 2\gamma\sqrt{[(D_{pq} + \gamma D_{qq})]^2 + (D_{qq}\omega_0^2)^2}]$$

*2.2. Mixing time: optimal exponential rate*

We show the preliminary calculations for high dimensions $d >> 1$. We consider $x \in \mathbb{R}^d$, $p \in \mathbb{R}^d$, so $(x,p) \in \mathbb{R}^{2d}$, and therefore the diffusion matrix is the block matrix

$$D = \begin{pmatrix} D_{qq}I_d & D_{pq}I_d \\ D_{pq}I_d & D_{pp}I_d \end{pmatrix}, \quad D_{pq} = D_{qp},$$

where $D_{ij} \in \mathbb{R}^{d \times d}$, $i,j = q,p$. As mentioned by Sparber et al. in [6], the optimal exponential rate $\kappa$ is the smalles eigenvalue of the Hessian of $\widetilde{A}$. The potential $\widetilde{A}$ is defined by

$$\widetilde{A}(\widetilde{y}) = A(\sqrt{D}y^T),$$

with $\sqrt{D}$ the square root of $D$ in the sense of positive definite matrices, and $y = (x,p)$.

We know the form of $A$ for the steady state, given by

$$A(x,p) = \frac{\gamma}{Q}[Q_{11}\omega_0^2 x^2 + 2Q_{12}\omega_0 x \cdot p + Q_{22}p^2]$$

$$Q_{11} = D_{pp} + \omega_0^2 D_{qq}$$

$$Q_{12} = 2\omega_0 \gamma D_{qq}$$

$$Q_{22} = D_{pp} + \omega_0^2 D_{qq} + 4\gamma(D_{pq} + \gamma D_{qq}) = Q_{11} + 4\gamma(D_{pq} + \gamma D_{qq})$$

$$Q = Q_{11}Q_{22} - Q_{12}^2$$

Now, if we consider

$$A(x,p) = (x,p)[\frac{\gamma}{Q}\begin{pmatrix} Q_{11}\omega_0^2 & Q_{12}\omega_0 \\ Q_{12}\omega_0 & Q_{22} \end{pmatrix}](x,p)^T$$

$$A(y^T) = y[\frac{\gamma}{Q}\begin{pmatrix} Q_{11}\omega_0^2 & Q_{12}\omega_0 \\ Q_{12}\omega_0 & Q_{22} \end{pmatrix}]y^T, \quad y = (x,p)$$

and

$$\widetilde{A}(\widetilde{y}) = A(\sqrt{D}y^T)$$

then, using the one square root $\sqrt{D}$ that is both positive semidefinite and symmetric, then

$$\widetilde{A}(\widetilde{y}) = \widetilde{y}[\frac{\gamma}{Q}\begin{pmatrix} Q_{11}\omega_0^2 & Q_{12}\omega_0 \\ Q_{12}\omega_0 & Q_{22} \end{pmatrix}]\widetilde{y}^T$$

$$\widetilde{A}(\widetilde{y}) = y\sqrt{D}[\frac{\gamma}{Q}\begin{pmatrix} Q_{11}\omega_0^2 & Q_{12}\omega_0 \\ Q_{12}\omega_0 & Q_{22} \end{pmatrix}]\sqrt{D}\widetilde{y}^T$$

$$A(\sqrt{D}y^T) = (x,p)[\frac{\gamma}{Q}\sqrt{D}\begin{pmatrix} Q_{11}\omega_0^2 & Q_{12}\omega_0 \\ Q_{12}\omega_0 & Q_{22} \end{pmatrix}\sqrt{D}](x,p)^T$$

and since we need

$$\frac{\partial^2 \widetilde{A}(\widetilde{y})}{\partial \widetilde{y}_i \widetilde{y}_j}, \quad i,j = 1, \cdots, 2d$$

to obtain the smallest eigenvalue of the Hessian of $\widetilde{A}$ under the transformed coordinates, we can just compute

$$\frac{\partial^2 A(y)}{\partial y_i y_j}$$

as it will render the same Hessian and eigenvalue under the previous coordinates.
So we have

$$H = \text{Hessian}(A) = \frac{\gamma}{Q}\begin{pmatrix} Q_{11}\omega_0^2 I_d & Q_{12}\omega_0 I_d \\ Q_{12}\omega_0 I_d & Q_{22} I_d \end{pmatrix}$$

with

$$Q_{11} = D_{pp} + \omega_0^2 D_{qq}$$
$$Q_{12} = 2\omega_0 \gamma D_{qq}$$
$$Q_{22} = D_{pp} + \omega_0^2 D_{qq} + 4\gamma(D_{pq} + \gamma D_{qq}) = Q_{11} + 4\gamma(D_{pq} + \gamma D_{qq})$$
$$Q = Q_{11}Q_{22} - Q_{12}^2$$

To find the smallest eigenvalue of the Hessian, we would need to solve the characteristic polynomial. Though this is manageable for the usual physical dimensions $1 \leq d \leq 3$, for high dimensions this might be unfeasible unless the structure of $H$ is simple. Therefore, let's study the case

$$0 = Q_{12} = 2\omega_0 \gamma D_{qq}$$

and since we don't want to make $\gamma = 0$ since the steady state would be trivial, neither $\omega_0 = 0$ since the harmonic potential would vanish, let's choose

$$D_{qq} = 0$$

which is somehow equivalent to a Caldeira-Legget model for $\gamma > 0$. Though this is unphysical, since the Caldeira-Legget model is known to not satisfy the Lindblad condition $D_{pp}D_{qq} - D_{pq}^2 = \det(D) \geq (\gamma/2)^2$, for the sake of simplicity we will proceed to perform its theoretical analysis. Then we have in that case

$$Q_{11} = D_{pp}$$

$$Q_{12} = 0$$
$$Q_{22} = D_{pp} + 4\gamma D_{pq}$$
$$Q = Q_{11}Q_{22} = D_{pp}^2 + 4\gamma D_{pq}D_{pp}$$

but in that case

$$H = \text{Hessian}(A) = \frac{\gamma}{Q}\begin{pmatrix} Q_{11}\omega_0^2 I_d & 0I_d \\ 0I_d & Q_{22}I_d \end{pmatrix}$$

$$H = \frac{\gamma}{Q}\begin{pmatrix} D_{pp}\omega_0^2 I_d & 0I_d \\ 0I_d & (D_{pp}^2 + 4\gamma D_{pq}D_{pp})I_d \end{pmatrix}$$

The eigenvalues of $H$ are clearly

$$\lambda_1 = \frac{\gamma}{Q}D_{pp}\omega_0^2$$

$$\lambda_2 = \frac{\gamma}{Q}(D_{pp}^2 + 4\gamma D_{pq}D_{pp})$$

Though we still have to determine the smallest root, it is clear that fact will be independent of the dimension $d$ of the problem. So we proceed to compare by looking for the case in which they are equal

$$0 = \frac{\gamma}{Q}D_{pp}(D_{pp} + 4\gamma D_{pq} - \omega_0^2)$$

and avoiding the trivial case $D_{pp} = 0$ (since then the eigenvalues are equally zero, so there would be no exponential convergence), then which is the smallest root is determined by

$$D_{pp} + 4\gamma D_{pq} \quad \text{vs.} \quad \omega_0^2$$

depending on which case one would have

$$\kappa = \lambda_i, \quad i = 1 \quad \text{or} \quad 2.$$

2.3. *Optimal exponential rate beyond Caldeira-Legget: case $Q_{11} = Q_{22}$*

Though formally we have

$$H = \text{Hessian}(A) = \frac{\gamma}{Q}\begin{pmatrix} Q_{11}\omega_0^2 I_d & Q_{12}\omega_0 I_d \\ Q_{12}\omega_0 I_d & Q_{22}I_d \end{pmatrix}$$

with

$$Q_{11} = D_{pp} + \omega_0^2 D_{qq}$$
$$Q_{12} = 2\omega_0\gamma D_{qq}$$
$$Q_{22} = D_{pp} + \omega_0^2 D_{qq} + 4\gamma(D_{pq} + \gamma D_{qq}) = Q_{11} + 4\gamma(D_{pq} + \gamma D_{qq})$$
$$Q = Q_{11}Q_{22} - Q_{12}^2$$

the particular case with $Q_{11} = Q_{22}$, which entails ($\gamma = 0$ is discarded as that trivial case makes the polynomial in the Gaussian vanish)

$$0 = 4\gamma(D_{pq} + \gamma D_{qq}) \implies D_{pq} = -\gamma D_{qq}$$

$$D = \begin{pmatrix} D_{qq} & -\gamma D_{qq} \\ -\gamma D_{qq} & D_{pp} \end{pmatrix} = D_{qq}\begin{pmatrix} 1 & -\gamma \\ -\gamma & D_{pp}/D_{qq} \end{pmatrix}$$

would render a very simple matrix

$$H = \frac{\gamma}{Q}\begin{pmatrix} Q_{11}\omega_0^2 I_d & Q_{12}\omega_0 I_d \\ Q_{12}\omega_0 I_d & Q_{11} I_d \end{pmatrix}$$

in which the transformation of coordinates (which amounts to some sort of linear combination of position and momentum operators: in properly normalized coordinates, is mixing the position variables with their respective canonically conjugated momentum variables)

$$w_j = y_j + y_{j+d}, \quad z_j = y_j - y_{j+d}, \quad j = 1, \cdots, d$$

would render the matrix representation of $H$ in the new basis as

$$[H]_{\{w,z\}} = \frac{\gamma}{Q}\begin{pmatrix} [Q_{11}\omega_0^2 + Q_{12}\omega_0]I_d & 0I_d \\ 0I_d & [Q_{11}\omega_0^2 - Q_{12}\omega_0]I_d \end{pmatrix}$$

$$[H]_{\{w,z\}} = \frac{\gamma}{Q}\omega_0\begin{pmatrix} [Q_{11}\omega_0 + Q_{12}]I_d & 0I_d \\ 0I_d & [Q_{11}\omega_0 - Q_{12}]I_d \end{pmatrix}$$

and in these new coordinates,

$$(w, z)^T = M^{-1}y^T$$

$$M^{-1} = \begin{pmatrix} I_d & I_d \\ I_d & -I_d \end{pmatrix}$$

$$y^T = M(w, z)^T$$

$$M = 2M^{-1} = 2\begin{pmatrix} I_d & I_d \\ I_d & -I_d \end{pmatrix}, \quad M = M^T$$

$$A(y^T) = y[\frac{\gamma}{Q}\begin{pmatrix} Q_{11}\omega_0^2 & Q_{12}\omega_0 \\ Q_{12}\omega_0 & Q_{22} \end{pmatrix}]y^T, \quad y = (x, p)$$

$$A(y^T) = (w, z)M[\frac{\gamma}{Q}\begin{pmatrix} Q_{11}\omega_0^2 & Q_{12}\omega_0 \\ Q_{12}\omega_0 & Q_{22} \end{pmatrix}]M(w, z)^T, \quad y = (x, p)$$

and

$$\tilde{A}(\tilde{y}) = A(\sqrt{D}y^T)$$

then, using the one square root $\sqrt{D}$ that is both positive semidefinite and symmetric, then

$$A(\sqrt{D}y^T) = (w, z)M\sqrt{D}[\frac{\gamma}{Q}\begin{pmatrix} Q_{11}\omega_0^2 & Q_{12}\omega_0 \\ Q_{12}\omega_0 & Q_{22} \end{pmatrix}]\sqrt{D}M(w, z)^T, \quad y = (x, p)$$

$$\widetilde{A}(\widetilde{y}) = \widetilde{y}[\frac{\gamma}{Q}(\begin{matrix} Q_{11}\omega_0^2 & Q_{12}\omega_0 \\ Q_{12}\omega_0 & Q_{22} \end{matrix})]\widetilde{y}^T$$

$$\widetilde{A}(\widetilde{y}) = y\sqrt{D}[\frac{\gamma}{Q}(\begin{matrix} Q_{11}\omega_0^2 & Q_{12}\omega_0 \\ Q_{12}\omega_0 & Q_{22} \end{matrix})]\sqrt{D}\widetilde{y}^T$$

$$A(\sqrt{D}y^T) = (x,p)[\frac{\gamma}{Q}\sqrt{D}(\begin{matrix} Q_{11}\omega_0^2 & Q_{12}\omega_0 \\ Q_{12}\omega_0 & Q_{22} \end{matrix})\sqrt{D}](x,p)^T$$

and since we need

$$\frac{\partial^2 \widetilde{A}(\widetilde{y})}{\partial \widetilde{y}_i \widetilde{y}_j}, \quad i,j = 1, \cdots, 2d$$

to obtain the smallest eigenvalue of the Hessian of $\widetilde{A}$ under the transformed coordinates, we can just compute

$$\frac{\partial^2 A(y)}{\partial y_i y_j}$$

as it will render the same Hessian and eigenvalue under the previous coordinates.
So we have

$$H = \text{Hessian}(A) = \frac{\gamma}{Q}(\begin{matrix} Q_{11}\omega_0^2 I_d & Q_{12}\omega_0 I_d \\ Q_{12}\omega_0 I_d & Q_{11} I_d \end{matrix})$$

which by transformation of coordinates $y \to (w,z)$ we can reorganize as

$$H = \text{Hessian}(A) = \frac{\gamma}{Q}(\begin{matrix} [Q_{11}\omega_0^2 + Q_{12}\omega_0]I_d & 0_d \\ 0_d & [Q_{11}\omega_0^2 - Q_{12}\omega_0]I_d \end{matrix})$$

$$H = \text{Hessian}(A) = \frac{\gamma}{Q}(\begin{matrix} [Q_{11}\omega_0 + Q_{12}]\omega_0 I_d & 0_d \\ 0_d & [Q_{11}\omega_0 - Q_{12}]\omega_0 I_d \end{matrix})$$

and in which case the characteristic polynomial of this Hessian would be

$$[Q_{11}\omega_0^2 + Q_{12}\omega_0 - \lambda][[Q_{11}\omega_0^2 - Q_{12}\omega_0 - \lambda] = 0$$

$$(\lambda - Q_{11}\omega_0^2 - Q_{12}\omega_0)(\lambda - Q_{11}\omega_0^2 + Q_{12}\omega_0) = 0$$

so the optimal exponential convergence rate (mixing time) is the smallest of the two possible roots

$$\lambda_1 = \frac{\gamma}{Q}[Q_{11}\omega_0 + Q_{12}]\omega_0$$

$$\lambda_2 = \frac{\gamma}{Q}[Q_{11}\omega_0 - Q_{12}]\omega_0$$

and since

$$Q_{11} = D_{pp} + \omega_0^2 D_{qq}$$

$$Q_{12} = 2\omega_0 \gamma D_{qq}$$

then

$$\lambda_1 = \frac{\gamma}{Q}[(D_{pp} + \omega_0^2 D_{qq})\omega_0 + 2\omega_0 \gamma D_{qq}]\omega_0$$

$$\lambda_2 = \frac{\gamma}{Q}[(D_{pp} + \omega_0^2 D_{qq})\omega_0 - 2\omega_0\gamma D_{qq}]\omega_0$$

so

$$\lambda_1 = \frac{\gamma}{Q}[D_{pp} + (\omega_0^2 + 2\gamma)D_{qq}]\omega_0^2$$

$$\lambda_2 = \frac{\gamma}{Q}[D_{pp} + (\omega_0^2 - 2\gamma)D_{qq}]\omega_0^2$$

therefore in this special case $Q_{11} = Q_{22} \iff D_{pq} = -\gamma D_{qq}$ we have that the optimal exponential convergence rate to the steady state is

$$\kappa = \frac{\gamma}{Q}[D_{pp} + (\omega_0^2 \pm 2\gamma)D_{qq}]\omega_0^2$$

which is also independent of the dimension $d$ of the problem in this special case.

*2.4. General case*

*2.4.1. General case for Hessian without perturbative approximations.*

To consider the general case of

$$H = \mathrm{Hessian}(A) = \frac{\gamma}{Q}\begin{pmatrix} Q_{11}\omega_0^2 I_d & Q_{12}\omega_0 I_d \\ Q_{12}\omega_0 I_d & Q_{22} I_d \end{pmatrix}$$

one can always do a change of variables (re-scaling mostly), such that, for $z = Ly$ (where $L$ is not only linear but diagonal and amounting to rescale each one of the respective coordinates), one could have (in fact part of the rescaling could help to make $\omega_0 = 1$)

$$H = \mathrm{Hessian}(A) = \frac{\gamma}{Q}\begin{pmatrix} q_{11} I_d & q_{12} I_d \\ q_{12} I_d & q_{11} I_d \end{pmatrix}$$

in which case

$$\kappa = \frac{\gamma}{Q}(q_{11} \pm q_{12})$$

$$\kappa = \frac{\gamma}{Q}[D_{pp} + (1 \pm 2\gamma)D_{qq}]$$

which is independent of the dimension $d$ of the vector variables. In fact, the rescaling can be performed in the momentum variables $p_j = y_{d+j}, \quad j = 1, \cdots, d$. By choosing $L = diag(1, \cdots, 1, Q_{11}/Q_{22}, \cdots, Q_{11}/Q_{22})$, one would get something equivalent to $Q_{11} = Q_{22}$ from the beginning, as in $q_{11} = q_{22}$, and then this particular case could handle the general one. In fact, let's go back to the expression of the Gaussian steady state in the Wigner formulation for WFP under a harmonic potential:

$$A(x,p) = \frac{\gamma}{Q}[Q_{11}\omega_0^2 x^2 + 2Q_{12}\omega_0 x \cdot p + Q_{22}p^2]$$

$$Q_{11} = D_{pp} + \omega_0^2 D_{qq}$$

$$Q_{12} = 2\omega_0\gamma D_{qq}$$

$$Q_{22} = D_{pp} + \omega_0^2 D_{qq} + 4\gamma(D_{pq} + \gamma D_{qq}) = Q_{11} + 4\gamma(D_{pq} + \gamma D_{qq})$$
$$Q = Q_{11}Q_{22} - Q_{12}^2$$

Let's consider the change of variables

$$x \to \omega_0 x$$

in which case we have

$$A(\omega_0 x, p) = \frac{\gamma}{Q}[Q_{11}(\omega_0 x)^2 + 2Q_{12}(\omega_0 x) \cdot p + Q_{22}p^2]$$

which is equivalent to take $\omega_0 = 1$, in which case

$$A(x, p) = \frac{\gamma}{Q}[Q_{11}x^2 + 2Q_{12}x \cdot p + Q_{22}p^2]$$

Let's assume for the moment the ratio $Q_{22}/Q_{11}$ is non-negative (meaning that these two numbers don't have opposite sign). In that case,

$$A(x, p) = \frac{\gamma Q_{11}}{Q}[x^2 + 2\frac{Q_{12}}{Q_{11}}x \cdot p + \frac{Q_{22}}{Q_{11}}p^2]$$

and, under the assumption $Q_{11}$ and $Q_{22}$ don't have opposite sign, we propose the transformation

$$p \to \sqrt{\frac{Q_{22}}{Q_{11}}}p$$

in which case we have

$$A(x, \sqrt{\frac{Q_{22}}{Q_{11}}}p) = \frac{\gamma Q_{11}}{Q}[x^2 + 2\frac{Q_{12}}{Q_{11}}\sqrt{\frac{Q_{11}}{Q_{22}}}x \cdot \sqrt{\frac{Q_{22}}{Q_{11}}}p + \frac{Q_{22}}{Q_{11}}p^2]$$

$$A(x, \sqrt{\frac{Q_{22}}{Q_{11}}}p) = \frac{\gamma Q_{11}}{Q}[x^2 + 2\frac{Q_{12}}{\sqrt{Q_{11}Q_{22}}}x \cdot \sqrt{\frac{Q_{22}}{Q_{11}}}p + (\sqrt{\frac{Q_{22}}{Q_{11}}}p)^2]$$

and, under these variables, the Hessian is

$$[H]_{\{x, \sqrt{Q_{22}}p/\sqrt{Q_{11}}\}} = \frac{\gamma Q_{11}}{Q}\begin{pmatrix} 1 & \frac{Q_{12}}{\sqrt{Q_{11}Q_{22}}} \\ \frac{Q_{12}}{\sqrt{Q_{11}Q_{22}}} & 1 \end{pmatrix}$$

$$[H]_{\{x, \sqrt{Q_{22}}p/\sqrt{Q_{11}}\}} = \frac{\gamma}{Q}\begin{pmatrix} Q_{11} & \frac{Q_{12}\sqrt{Q_{11}}}{\sqrt{Q_{22}}} \\ \frac{Q_{12}\sqrt{Q_{11}}}{\sqrt{Q_{22}}} & Q_{11} \end{pmatrix}$$

so, under the proposed transformation of variables,

$$(x, p) \to (\omega_0 x, \sqrt{\frac{Q_{22}}{Q_{11}}}p)$$

the Hessian is

$$[H]_{\{x,\sqrt{\frac{Q_{22}}{Q_{11}}}p\}} = \frac{\gamma}{Q}\begin{pmatrix} Q_{11} & Q_{12}\sqrt{\frac{Q_{11}}{Q_{22}}} \\ Q_{12}\sqrt{\frac{Q_{11}}{Q_{22}}} & Q_{11} \end{pmatrix}$$

in which now the transformation of coordinates of linear combination can render a Hessian with diagonal structure: if we propose

$$(\omega_0 x, \sqrt{\frac{Q_{22}}{Q_{11}}}p) \to (\omega_0 x + \sqrt{\frac{Q_{22}}{Q_{11}}}p, \omega_0 x - \sqrt{\frac{Q_{22}}{Q_{11}}}p)$$

the Hessian will have the diagonal structure

$$[H]_{\{\omega_0 x + \sqrt{\frac{Q_{22}}{Q_{11}}}p, \omega_0 x - \sqrt{\frac{Q_{22}}{Q_{11}}}p\}} = \frac{\gamma}{Q}\begin{pmatrix} Q_{11} + Q_{12}\sqrt{\frac{Q_{11}}{Q_{22}}} & 0_d \\ 0_d & Q_{11} - Q_{12}\sqrt{\frac{Q_{11}}{Q_{22}}} \end{pmatrix}$$

and under this change of variables, the eigenvalues of the Hessian are only two, namely

$$\kappa_\pm = \frac{\gamma}{Q}(Q_{11} \pm Q_{12}\sqrt{\frac{Q_{11}}{Q_{22}}})$$

and the optimal convergence rate is the smallest of them. To determine which one is the smallest, we just have to compare them. So we have

$$\kappa_\pm = \frac{\gamma\sqrt{Q_{11}}}{Q}(\sqrt{Q_{11}} \pm \frac{Q_{12}}{\sqrt{Q_{22}}})$$

and recalling

$$Q_{11} = D_{pp} + \omega_0^2 D_{qq}$$

$$Q_{12} = 2\omega_0\gamma D_{qq}$$

$$Q_{22} = D_{pp} + \omega_0^2 D_{qq} + 4\gamma(D_{pq} + \gamma D_{qq}) = Q_{11} + 4\gamma(D_{pq} + \gamma D_{qq})$$

$$Q = Q_{11}Q_{22} - Q_{12}^2$$

the optimal convergence rate is the smallest of the following two roots:

$$\kappa_\pm = \frac{\gamma}{Q_{11}Q_{22} - Q_{12}^2}(D_{pp} + \omega_0^2 D_{qq} \pm \frac{2\omega_0\gamma D_{qq}}{\sqrt{1 + 4\gamma\frac{D_{pq}+\gamma D_{qq}}{Q_{11}}}})$$

$$\kappa_\pm = \frac{\gamma}{Q_{11}[Q_{11} + 4\gamma(D_{pq} + \gamma D_{qq})] - [2\omega_0\gamma D_{qq}]^2}(D_{pp}+\omega_0^2 D_{qq}\pm\frac{1}{\sqrt{\frac{1}{(2\omega_0\gamma D_{qq})^2} + \frac{4\gamma}{(2\omega_0\gamma D_{qq})^2}\frac{D_{pq}+\gamma D_{qq}}{Q_{11}}}})$$

$$\kappa_\pm = \frac{\gamma}{Q_{11}^2 + 4\gamma(D_{pq} + \gamma D_{qq})Q_{11} - 4\omega_0^2\gamma^2 D_{qq}^2}(D_{pp}+\omega_0^2 D_{qq}\pm\frac{1}{\sqrt{\frac{1}{(2\omega_0\gamma D_{qq})^2} + \frac{1}{(\omega_0 D_{qq})^2\gamma}\frac{D_{pq}+\gamma D_{qq}}{Q_{11}}}})$$

and, under the rescaling that makes $\omega_0 = 1$, we have

$$\kappa_\pm = \frac{\gamma}{Q_{11}^2 + 4\gamma(D_{pq} + \gamma D_{qq})Q_{11} - 4\gamma^2 D_{qq}^2}(D_{pp}+D_{qq}\pm\frac{1}{\sqrt{\frac{1}{(2\gamma D_{qq})^2} + \frac{1}{D_{qq}^2\gamma}\frac{D_{pq}+\gamma D_{qq}}{Q_{11}}}})$$

$$\kappa_\pm = \frac{\gamma}{(D_{pp}+D_{qq})^2 + 4\gamma(D_{pq}+\gamma D_{qq})(D_{pp}+D_{qq}) - 4\gamma^2 D_{qq}^2}(D_{pp}+D_{qq}\pm\frac{1}{\sqrt{\frac{1}{(2\gamma D_{qq})^2} + \frac{1}{D_{qq}^2\gamma}\frac{D_{pq}+\gamma D_{qq}}{D_{pp}+D_{qq}}}})$$

$$\kappa_\pm = \frac{\gamma}{(D_{pp}+D_{qq})^2 + (4\gamma D_{pq} + 4\gamma^2 D_{qq})(D_{pp}+D_{qq}) - 4\gamma^2 D_{qq}^2}(D_{pp}+D_{qq}\pm\frac{1}{\sqrt{\frac{1}{(2\gamma D_{qq})^2} + \frac{1}{D_{qq}^2\gamma}\frac{D_{pq}+\gamma D_{qq}}{D_{pp}+D_{qq}}}})$$

$$\kappa_\pm = \frac{\gamma}{(D_{pp}+D_{qq})^2 + 4\gamma D_{pq}(D_{pp}+D_{qq}) + 4\gamma^2 D_{qq}(D_{pp}+D_{qq}) - 4\gamma^2 D_{qq}^2}(D_{pp}+D_{qq}\pm\frac{1}{\sqrt{\frac{1}{(2\gamma D_{qq})^2} + \frac{1}{D_{qq}^2\gamma}\frac{D_{pq}+\gamma D_{qq}}{D_{pp}+D_{qq}}}})$$

$$\kappa_\pm = \frac{\gamma}{(D_{pp}+D_{qq})^2 + 4\gamma D_{pq}(D_{pp}+D_{qq}) + 4\gamma^2 D_{qq}D_{pp}}(D_{pp}+D_{qq}\pm\frac{1}{\sqrt{\frac{1}{(2\gamma D_{qq})^2} + \frac{1}{D_{qq}}\frac{1+D_{pq}/D_{qq}\gamma}{D_{pp}+D_{qq}}}})$$

$$\kappa_\pm = \frac{\gamma}{(D_{pp}+D_{qq})^2 + 4\gamma D_{pq}(D_{pp}+D_{qq}) + 4\gamma^2 D_{qq}D_{pp}}(D_{pp}+D_{qq}\pm[\frac{1}{(2\gamma D_{qq})^2} + \frac{1}{D_{qq}}\frac{1+\frac{D_{pq}}{\gamma D_{qq}}}{D_{pp}+D_{qq}}]^{-1/2})$$

$$\kappa_\pm = \frac{\gamma}{D_{pp}^2 + D_{qq}^2 + 2(1+2\gamma^2)D_{qq}D_{pp} + 4\gamma D_{pq}(D_{pp}+D_{qq})}(D_{pp}+D_{qq}\pm[\frac{1}{(2\gamma D_{qq})^2} + \frac{1}{D_{qq}}\frac{1+\frac{D_{pq}}{\gamma D_{qq}}}{D_{pp}+D_{qq}}]^{-1/2})$$

$$\kappa_\pm = \frac{\gamma}{(1+2\gamma^2)(D_{pp}+D_{qq})^2 - 4\gamma^2 D_{qq}D_{pp} + 4\gamma D_{pq}(D_{pp}+D_{qq})}(D_{pp}+D_{qq}\pm[\frac{1}{(2\gamma D_{qq})^2} + \frac{1}{D_{qq}}\frac{1+\frac{D_{pq}}{\gamma D_{qq}}}{D_{pp}+D_{qq}}]^{-1/2})$$

$$\kappa_\pm = \frac{D_{pp}+D_{qq}\pm[\frac{1}{(2\gamma D_{qq})^2} + (1+\frac{D_{pq}}{\gamma D_{qq}})\frac{1}{D_{qq}(D_{pp}+D_{qq})}]^{-1/2}}{(\frac{1}{\gamma}+2\gamma)(D_{pp}+D_{qq})^2 - 4\gamma D_{qq}D_{pp} + 4D_{pq}(D_{pp}+D_{qq})}$$

$$\kappa_\pm = \frac{D_{pp}+D_{qq}\pm 1/\sqrt{\frac{1}{(2\gamma D_{qq})^2} + (1+\frac{D_{pq}}{\gamma D_{qq}})\frac{1}{D_{qq}(D_{pp}+D_{qq})}}}{(\frac{1}{\gamma}+2\gamma)(D_{pp}+D_{qq})^2 + 4[D_{pq}(D_{pp}+D_{qq}) - \gamma D_{qq}D_{pp}]}$$

$$\kappa_\pm = \frac{D_{pp}+D_{qq}\pm 1/\sqrt{\frac{1}{D_{qq}^2}[\frac{1}{(2\gamma)^2} + \frac{1+D_{pq}/D_{qq}\gamma}{1+D_{pp}/D_{qq}}]}}{(\frac{1}{\gamma}+2\gamma)(D_{pp}+D_{qq})^2 + 4[D_{pq}(D_{pp}+D_{qq}) - \gamma D_{qq}D_{pp}]}$$

## 3. Stochastic Gradient Descent in the continuous-time limit as an SDE: Classical vs. quantum regimes

We focus on the study of stochastic gradient descent by means of their continuous time limit (Su, Wibisono, Jordan, Shi).

We can understand the Stochastic Gradient Descent (SGD) algorithm by following [Liu, Su, Li], in which they consider $f : \mathbb{R}^d \to \mathbb{R}^d$ the objective function constructed with all data, and the SGD algorithm being given by

$$x_{k+1} = x_k - s\widetilde{\nabla} f(x_k)$$

where $s$ is the so-called "learning rate" and $\widetilde{\nabla} f$ is the estimated gradient obtained from a mini-batch. One can model SGD then by the following iterative method,

$$x_{k+1} = x_k - s\nabla f(x_k) - s\xi_k$$

where $\xi_k$ is a given noise that obeys a normal distribution, as in the so-called "unadjusted Langevin dynamics". Taking the continuous time limit one obtains a stochastic differential equation (SDE) that includes the learning rate $s$ as a parameter, which models the discrete SGD algorithm. This SDE is

$$dx = -\nabla f(x)dt + \sqrt{s}dW,$$

with $W$ a Brownian motion. It is indicating that the transport is such that the objective function is minimized by following the direction opposite to the gradient, namely $-\nabla f$, but being also subjected to a noise following a normal distribution.

On the other hand, the main equation in quantum mechanics where both transport and noise are combined is a master equation, either in a Wigner formulation, such as Wigner-Fokker-Planck (integro-differential equation in continuous variables)

$$w_t + p \cdot w_x + \Theta[V](w) = \nabla \cdot (D\nabla w) + 2\gamma \nabla_p \cdot (wp)$$

or a density matrix one, such as the Lindblad master equation (an operator equation, which can be projected in any basis such as a position one),

$$\rho_t = -\frac{i}{\hbar}[H, \rho] + \sum_i \gamma_i (L_i \rho L_i^\dagger - \frac{1}{2}\{L_i L_i^\dagger, \rho\})$$

and there is a third way to represent this noisy quantum transport process in an SDE form via a "Quantum Langevin" equation, where defining $z = (x, p)$ as a phase-space position-momentum vector, is given by (for the case of a Harmonic potential)

$$dz = Q[z(t)]dt + E[z_i(t)]$$

with $Q[z(t)] = (p, -x - 2\gamma p)$ being the quantum transport, and $E$ being a zero average Gaussian random variable, whose variance is $Ddt$.

The question is: are our quantum master equations above quantum analogs of the SDE equation with a gradient transport equivalent to a continuous time limit of SGD?

Let's recall the classical limit of the WFP equation, as indicated in [6]. We have that [6] in the classical limit $\hbar \to 0$, $D_{qq} = 0 = D_{pq}$ and the pseudo-differential operator becomes in this limit one of the phase-space Liouvillean transports, as in (taking $m = 1$)

$$\Theta[V]\{w\} = -\nabla_x V \cdot \nabla_p w$$

so the classical Fokker-Planck equation in phase space for the probability distribution $f$ is recovered as $w \to f$ when $\hbar \to 0$, namely

$$f_t + p \cdot f_x - \nabla_x V \cdot f_p = \nabla_p \cdot (D_{pp} \nabla_p f) + 2\gamma \nabla_p \cdot (fp)$$

and since we have the Hamiltonian

$$H(x,p) = \frac{p^2}{2} + V(x)$$

part of the transport has a symplectic structure, as in

$$f_t + \nabla_p H \cdot f_x - \nabla_x H \cdot f_p = \nabla_p \cdot (D_{pp} \nabla_p f) + 2\gamma \nabla_p \cdot (fp)$$

and though we could express this as Poisson brackets, we can express the Hamiltonian transport as

$$f_t + \nabla \cdot (f(\nabla_p H, -\nabla_x H)) = \nabla_p \cdot (D_{pp} \nabla_p f) + 2\gamma \nabla_p \cdot (fp)$$

and adding the friction to the Hamiltonian transport we get the classical kinetic Fokker-Planck equation as

$$f_t + \nabla \cdot (f(\nabla_p H, -\nabla_x H)) - 2\gamma \nabla \cdot (f(0,p)) = \nabla_p \cdot (D_{pp} \nabla_p f)$$

$$f_t + \nabla \cdot (f(\nabla_p H, -\nabla_x H - 2\gamma p)) = \nabla_p \cdot (D_{pp} \nabla_p f)$$

so again our transport is

$$\alpha = (p, -\nabla_x V - 2\gamma p)$$

so for example, if $V = x^2/2$ is harmonic, we have

$$\alpha = (p, -x - 2\gamma p)$$

The problem is that a classical kinetic Fokker-Planck is just Hamiltonian transport and noise (with both diffusion and friction). On the other hand, the SDE for the continuous-time limit SGD is a gradient transport and noise (represented via diffusion). Namely, the SDE for the limit behavior of SGD has a steepest descent transport following the gradient with some added noise, whereas a Fokker-Planck equation has a Hamiltonian transport (the gradient multiplied by a symplectic matrix) with momentum noise and friction (the latter being the non-Hamiltonian contribution of the transport, rather in gradient form instead). The difference is clearly noticed if one thinks of the harmonic potential $V = x^2/2$, in which the equation without noise will just represent the harmonic oscillator orbiting the origin in phase space conserving the energy functional (on an ellipse or circle depending the scaling of coordinates), whereas the transport in the SGD goes in the opposite direction of the gradient for the objective function, minimizing the respective functional:

$$dx = -\nabla F(x)dt + \sqrt{s}dW,$$

and even the kinetic diffusive PDE representing the SGD in continuous time limit is known by Jordan et al. [3], as in

$$\partial_t \rho_{SGD} = \nabla \cdot (\rho_{SGD} \nabla F) + \frac{s}{2} \nabla^2 \rho_{SGD}$$

so the analog of a diffusion coefficient is $s/2$. Formally the variable in the equation above needs only be the position $x$: in the quantum version of SGD the problem dimensionality needs to be doubled by introducing a conjugate variable such as the momentum $p$. On the other hand, if we want to make analogies with the classical case it might be useful to think of a phase space setting even from the classical case, so we would have

$$\partial_t \rho_{SGD} = \nabla_{(x,p)} \cdot (\rho_{SGD} \nabla_{(x,p)} F) + \nabla_{(x,p)} \cdot (\frac{s}{2} I_{2d} \nabla_{(x,p)} \rho_{SGD})$$

and if $x, p \in \mathbb{R}^d$ then $D = sI_{2d}/2$ in this case. If this Fokker-Planck for the limit SGD has an analog with the physics-motivated classical Fokker-Planck

$$f_t + \nabla \cdot (f(p, -x - 2\gamma p)) = \nabla \cdot (D_{pp} \begin{pmatrix} 0_d & 0_d \\ 0_d & I_d \end{pmatrix} \nabla f)$$

then we need that

$$\nabla_{(x,p)} F = 2\gamma p, \quad \gamma >> 1$$

so the analog is strong when the objective function is, for $\gamma >> 1$,

$$F = \gamma p^2 + C$$

with $C$ a constant of integration to be defined. The analogy is not 100% direct, since $D_{qq} = 0$ yet for the SGD-SDE limit equation one has $D_{qq} = s/2$. This would seem to indicate that the quantum analog of an SGD can be found for quadratic functionals dominant over the other noise term (diffusion) and with a negligible Hamiltonian transport by taking $\gamma >> 1$.

**Appendix**

*Asymptotic approximation to the optimal convergence rate (Perturbation theory approach)*

We consider the general case of

$$H = \text{Hessian}(A) = \frac{\gamma}{Q} \begin{pmatrix} Q_{11} \omega_0^2 I_d & Q_{12} \omega_0 I_d \\ Q_{12} \omega_0 I_d & Q_{22} I_d \end{pmatrix}$$

and since

$$Q_{11} = D_{pp} + \omega_0^2 D_{qq}$$

$$Q_{12} = 2\omega_0 \gamma D_{qq}$$

$$Q_{22} = D_{pp} + \omega_0^2 D_{qq} + 4\gamma(D_{pq} + \gamma D_{qq}) = Q_{11} + 4\gamma(D_{pq} + \gamma D_{qq})$$

$$Q = Q_{11} Q_{22} - Q_{12}^2$$

First, by a convenient change of coordinates

$$x \to x/\omega_0$$

which is equivalent to assuming $\omega_0 = 1$, the Hessian matrix is rescaled to

$$H = \frac{\gamma}{Q}\begin{pmatrix} Q_{11}I_d & Q_{12}I_d \\ Q_{12}I_d & Q_{22}I_d \end{pmatrix}$$

which also renders the rescaling

$$Q_{11} = D_{pp} + D_{qq}$$

$$Q_{12} = 2\gamma D_{qq}$$

$$Q_{22} = D_{pp} + D_{qq} + 4\gamma(D_{pq} + \gamma D_{qq}) = Q_{11} + 4\gamma(D_{pq} + \gamma D_{qq})$$

Then we can split our matrix in a convenient way, as the sum of two matrices easier to analyze,

$$H = \frac{\gamma}{Q}\begin{pmatrix} Q_{11}I_d & Q_{12}I_d \\ Q_{12}I_d & [Q_{11} + 4\gamma(D_{pq} + \gamma D_{qq})]I_d \end{pmatrix}$$

$$H = \frac{\gamma}{Q}[Q_{11}\begin{pmatrix} I_d & 0_d \\ 0_d & I_d \end{pmatrix} + \begin{pmatrix} 0_d & Q_{12}I_d \\ Q_{12}I_d & 4\gamma(D_{pq} + \gamma D_{qq})I_d \end{pmatrix}]$$

$$H = \frac{\gamma}{Q}[Q_{11}I_{2d} + \begin{pmatrix} 0_d & Q_{12}I_d \\ Q_{12}I_d & 4\gamma(D_{pq} + \gamma D_{qq})I_d \end{pmatrix}]$$

If we assume

$$Q_{11} = D_{pp} + D_{qq} \neq 0$$

then

$$H = \frac{\gamma Q_{11}}{Q}[I_{2d} + \begin{pmatrix} 0_d & I_d Q_{12}/Q_{11} \\ I_d Q_{12}/Q_{11} & I_d 4\gamma(D_{pq} + \gamma D_{qq})/Q_{11} \end{pmatrix}]$$

and recalling

$$Q_{11} = D_{pp} + D_{qq}$$

$$Q_{12} = 2\gamma D_{qq}$$

then

$$H = \frac{\gamma Q_{11}}{Q}[I_{2d} + \begin{pmatrix} 0_d & I_d 2\gamma D_{qq}/(D_{pp} + D_{qq}) \\ I_d 2\gamma D_{qq}/(D_{pp} + D_{qq}) & I_d 4\gamma(D_{pq} + \gamma D_{qq})/(D_{pp} + D_{qq}) \end{pmatrix}]$$

$$H = \frac{\gamma Q_{11}}{Q}[I_{2d} + 2\gamma\begin{pmatrix} 0_d & I_d D_{qq}/(D_{pp} + D_{qq}) \\ I_d D_{qq}/(D_{pp} + D_{qq}) & I_d 2(D_{pq} + \gamma D_{qq})/(D_{pp} + D_{qq}) \end{pmatrix}]$$

assuming $D_{qq} \neq 0$, then

$$H = \frac{\gamma Q_{11}}{Q}[I_{2d} + 2\gamma\begin{pmatrix} 0_d & I_d/(1 + D_{pp}/D_{qq}) \\ I_d/(1 + D_{pp}/D_{qq}) & I_d 2\{1 + (\gamma - 1)/(1 + D_{pp}/D_{qq})\} \end{pmatrix}]$$

If we are able to obtain the eigenvalues of the second matrix, we can approximately obtain the eigenvalues of the full matrix by perturbation theory. Our matrix

$$\begin{pmatrix} 0_d & \frac{I_d}{(1+D_{pp}/D_{qq})} \\ \frac{I_d}{(1+D_{pp}/D_{qq})} & 2[1 + \frac{(\gamma-1)}{(1+D_{pp}/D_{qq})}]I_d \end{pmatrix}$$

which, if we define
$$r = \frac{1}{1 + D_{pp}/D_{qq}}$$
is given by
$$\begin{pmatrix} 0_d & rI_d \\ rI_d & 2[1 + (\gamma - 1)r]I_d \end{pmatrix}$$
We will use the following result.

**Lemma 2.** *If a real matrix $2d \times 2d$ has the structure*
$$T = \begin{pmatrix} 0_d & aI_d \\ aI_d & bI_d \end{pmatrix}$$
*under the assumption*
$$(a, b) \neq (0, 0)$$
*to avoid trivial cases, there are only two eigenvalues for $T$ of the form*
$$\frac{b}{2} \pm \sqrt{(b/2)^2 + a^2}$$

*Proof.* If we consider a vector of the form
$$\alpha e_j + \beta e_{j+d}, \quad j = 1, \cdots d$$
when applying the transformation above, we have that
$$T(\alpha e_j + \beta e_{j+d}) = \alpha a e_{j+d} + \beta(ae_j + be_{j+d}) = a\beta e_j + (\alpha a + \beta b)e_{j+d}$$
and an eigenvalue would be such that the following ratios are equal
$$\frac{a\beta}{\alpha} = \frac{\alpha a + \beta b}{\beta} = a\frac{\alpha}{\beta} + b$$
so
$$a\left(\frac{\beta}{\alpha} - \frac{\alpha}{\beta}\right) = b$$
equivalent to
$$a\left(\frac{\beta}{\alpha}\right)^2 - b\frac{\beta}{\alpha} - a = 0$$
and therefore the two possible solutions for the ratio $\beta/\alpha$ are
$$\left(\frac{\beta}{\alpha}\right)_\pm = \frac{b \pm \sqrt{b^2 + 4a^2}}{2a}$$
which are different under the assumption that $(a, b) \neq (0, 0)$, therefore the eigenvalues are of the form
$$\frac{a\beta}{\alpha} = \frac{b \pm \sqrt{b^2 + 4a^2}}{2} = \frac{b}{2} \pm \sqrt{\left(\frac{b}{2}\right)^2 + a^2}$$

since
$$T(\alpha e_j + \beta e_{j+d}) = \frac{a\beta}{\alpha}(\alpha e_j + \beta e_{j+d})$$

with respective eigenvectors
$$\alpha e_j + \beta e_{j+d} \propto e_j + (\frac{\beta}{\alpha})_{\pm} e_{j+d}, \quad j = 1, \cdots d$$

so we have found all our $2d$ possible eigenvalues (two cases $\pm$ for each $j = 1, \cdots, d$). Under this eigenbasis, our operator is represented by the diagonal matrix

$$[T]_{\{e_j + (\beta/\alpha)_+ e_{j+d}\}_\pm} = \begin{pmatrix} (\frac{b}{2} + \sqrt{(\frac{b}{2})^2 + a^2}) I_d & 0_d \\ 0_d & (\frac{b}{2} - \sqrt{(\frac{b}{2})^2 + a^2}) I_d \end{pmatrix}$$

□

By using the lemma above, if we consider our previous matrix

$$\begin{pmatrix} 0_d & rI_d \\ rI_d & 2[1 + (\gamma - 1)r]I_d \end{pmatrix}$$

then
$$a = r$$
$$b = 2[1 + (\gamma - 1)r]$$

so the eigenvalues of this matrix are
$$1 + (\gamma - 1)r \pm \sqrt{(1 + (\gamma - 1)r)^2 + r^2}$$

and going back to our previous example, let's consider the perturbative problem with parameter $\varepsilon = O(1/\gamma^2)$, so that $\gamma\varepsilon = O(1/\gamma) \ll 1$ since $\gamma \gg 1$,

$$I_{2d} + 2\gamma \begin{pmatrix} 0_d & rI_d \\ rI_d & 2[1 + (\gamma - 1)r]I_d \end{pmatrix} \varepsilon$$

Since the first matrix is the identity and the second one is symmetric (both real valued and $2d \times 2d$), one can prove [Holmes, Intro. to Perturbation Methods] that a two-term perturbative expansion of the eigenvalues $\mu_i$ of the matrix above are
$$\mu_i = 1 + \varepsilon \lambda_i, \quad i = 1, \cdots, 2d$$

with $\lambda_i$ the eigenvalues of the symmetric matrix. For the case $\varepsilon = 1$, we approximately have that the eigenvalues of

$$I_{2d} + 2\gamma \begin{pmatrix} 0_d & rI_d \\ rI_d & 2[1 + (\gamma - 1)r]I_d \end{pmatrix}$$

are
$$2 + (\gamma - 1)r \pm \sqrt{(1 + (\gamma - 1)r)^2 + r^2}$$

and if we finally go back to our original eigenproblem for the Hessian,

$$H = \frac{\gamma Q_{11}}{Q}[I_{2d} + 2\gamma(\begin{matrix} 0_d & I_d/(1 + D_{pp}/D_{qq}) \\ I_d/(1 + D_{pp}/D_{qq}) & I_d 2\{1 + (\gamma - 1)/(1 + D_{pp}/D_{qq})\} \end{matrix})]$$

remembering

$$r = \frac{1}{1 + D_{pp}/D_{qq}} = \frac{D_{qq}}{D_{pp} + D_{qq}}$$

then the eigenvalues of $H$, up to first order in a perturbative expansion, are only two and are namely

$$\frac{\gamma Q_{11}}{Q}[2 + (\gamma - 1)r \pm \sqrt{(1 + (\gamma - 1)r)^2 + r^2}]$$

and recalling

$$Q_{11} = D_{pp} + D_{qq}$$
$$Q_{12} = 2\gamma D_{qq}$$
$$Q_{11} + 4\gamma(D_{pq} + \gamma D_{qq})$$

then

$$\frac{\gamma}{Q}[2Q_{11} + (\gamma - 1)Q_{11}r \pm Q_{11}\sqrt{(1 + (\gamma - 1)r)^2 + r^2}]$$

$$\frac{\gamma}{Q}[2Q_{11} + (\gamma - 1)D_{qq} \pm \text{sign}(Q_{11})\sqrt{(Q_{11} + (\gamma - 1)Q_{11}r)^2 + (Q_{11}r)^2}]$$

$$\frac{\gamma}{Q}[2Q_{11} + (\gamma - 1)D_{qq} \pm \sqrt{(D_{pp} + D_{qq} + (\gamma - 1)D_{qq})^2 + D_{qq}^2}]$$

$$\frac{\gamma}{Q}[2(D_{pp} + D_{qq}) + (\gamma - 1)D_{qq} \pm \sqrt{(D_{pp} + D_{qq} + (\gamma - 1)D_{qq})^2 + D_{qq}^2}]$$

$$\frac{\gamma}{Q}[2D_{pp} + (\gamma + 1)D_{qq} \pm \sqrt{(D_{pp} + \gamma D_{qq})^2 + D_{qq}^2}]$$

and

$$Q = Q_{11}(Q_{11} + 4\gamma(D_{pq} + \gamma D_{qq})) - (2\gamma D_{qq})^2$$
$$Q = Q_{11}^2 + 4\gamma(D_{pq} + \gamma D_{qq})Q_{11} - (2\gamma D_{qq})^2$$
$$Q = (D_{pp} + D_{qq})^2 + 4\gamma(D_{pq} + \gamma D_{qq})(D_{pp} + D_{qq}) - (2\gamma D_{qq})^2$$

In conclusion, by perturbative methods, up to first order, the eigenvalues of the Hessian

$$H = \frac{\gamma Q_{11}}{Q}[I_{2d} + 2\gamma(\begin{matrix} 0_d & I_d/(1 + D_{pp}/D_{qq}) \\ I_d/(1 + D_{pp}/D_{qq}) & I_d 2\{1 + (\gamma - 1)/(1 + D_{pp}/D_{qq})\} \end{matrix})]$$

are only the following two:

$$\frac{\gamma[2D_{pp} + (\gamma + 1)D_{qq} \pm \sqrt{(D_{pp} + \gamma D_{qq})^2 + D_{qq}^2}]}{(D_{pp} + D_{qq})^2 + 4\gamma(D_{pq} + \gamma D_{qq})(D_{pp} + D_{qq}) - (2\gamma D_{qq})^2}$$

under the renormalization of the position variable $x$ equivalent to taking $\omega_0 = 1$.

Since the optimal exponential convergence rate $\kappa$ is the smallest eigenvalue of the Hessian above, then up to first order the convergence rate is

$$\kappa = \gamma \frac{2D_{pp} + (\gamma + 1)D_{qq} - \sqrt{(D_{pp} + \gamma D_{qq})^2 + D_{qq}^2}}{(D_{pp} + D_{qq})^2 + 4\gamma(D_{pq} + \gamma D_{qq})(D_{pp} + D_{qq}) - (2\gamma D_{qq})^2}$$

which (again, up to first order), is independent of the dimension $d$ of the phase-space variables $(x, p) \in \mathbb{R}^d \times \mathbb{R}^d$ of the problem.

*Remark* 3. The reason this perturbative calculation was kept as a preliminary calculation in the Appendix was because the perturbative argument splitting the matrix into an identity matrix and a remainder needs this remainder to be small (as for example with a perturbative parameter $\varepsilon >> 1$), however the remainder matrix is multiplied by the friction parameter $\gamma >> 1$ which we have assumed to be quite large in order to make the gradient transport due to the friction dominant over the Hamiltonian one. Therefore, though formally a correct calculation with the perturbative parameter $\varepsilon$, once removing $\varepsilon$ then the assumption needed for the perturbative calculation does not apply anymore and therefore we keep this perturbative analysis as a formal calculation for the Appendix only, but not the definite approach for our final results, presented previously in the bulk of the text.

———————————————————————————